\documentclass[12pt]{article}
\usepackage[T1]{fontenc}
\usepackage[utf8]{inputenc}

\usepackage{geometry}
\usepackage{amsmath}
\usepackage{amsthm}
\usepackage{amsfonts}
\usepackage{amssymb}
\usepackage{amsbsy}

\usepackage{cancel}
\usepackage{url}
\usepackage{mathdots}
\usepackage{systeme}
\usepackage{verbatim}
\usepackage{fancyhdr}
\usepackage{subcaption}
\usepackage{hyperref}
\usepackage{enumerate}
\usepackage{enumitem}
\usepackage{tikz}
\usepackage{graphicx}
\usepackage{color}
\usepackage{multicol}
\usepackage{tabu}
\usepackage{yhmath}

\usepackage{float}
\usepackage{mathtools}
\usepackage{titlesec}
\usepackage{wasysym}
\usepackage{titling}
\usepackage{xlop}
\usepackage{subfiles}
\usepackage{appendix}
\usepackage{csquotes}
\usepackage{textgreek}
\usepackage{epigraph}
\usepackage{minitoc}
\usepackage{latexsym}
\usepackage{newlfont}

\usepackage{hyperref}
\usepackage[capitalize,nameinlink,noabbrev]{cleveref}

\theoremstyle{definition}
\newtheorem{theorem}{Theorem}
\newtheorem{lemma}[theorem]{Lemma}

\newtheorem{example}[theorem]{Example}
\newtheorem{definition}[theorem]{Definition}

\usepackage{answers, xpatch}

\Newassociation{probsol}{Solution}{ans}

\theoremstyle{remark}

\begin{document}

\everymath{\displaystyle}

\begin{center}
    {\Large Functions with Diffusive Properties}\\
    Samer Seraj\\
    Existsforall Academy\\
    \href{mailto:samer\_seraj@outlook.com}{samer\_seraj@outlook.com}\\
    \today
\end{center}

\begin{abstract}
    While exploring desirable properties of hash functions in cryptography, the author was led to investigate three notions of functions with scattering or ``diffusive'' properties, where the functions map between binary strings of fixed finite length. These notions of diffusion ask for some property to be fulfilled by the Hamming distances between outputs corresponding to pairs of inputs that lie on the endpoints of edges of an $n$-dimensional hypercube. Given the dimension of the input space, we explicitly construct such functions for every dimension of the output space that allows for the functions to exist.
\end{abstract}

\bigskip

We begin by reminding the reader of some basic notions so that the utilized terminology, concepts and notation are clear.

\begin{definition}
Let $n$ be a positive integer.
\begin{enumerate}
	\item The set of \textbf{$n$-bit binary strings} are $n$-tuples of elements from the field $\{0,1\}$, denoted by $\{0,1\}^n$. The bits of each $x\in\{0,1\}^n$ are indexed from $1$ to $n$, from left to right. 
	\item The \textbf{XOR} binary operation $\oplus$ on $\{0,1\}^n$ is defined as bit-wise ``addition'' in the field $\{0,1\}$. A generalization to arbitrary pairs of finite bit strings is that sufficiently many $0$'s are first appended to the right end of the shorter string to force both strings to have the same length.
	\item The \textbf{Hamming weight} $w:\{0,1\}^n\to\mathbb{Z}$ of $x\in\{0,1\}^n$ is the number of non-zero bits of $x$.
	\item The \textbf{Hamming distance} $h:\{0,1\}^n\times\{0,1\}^n\to\mathbb{Z}$ between $x,y\in\{0,1\}^n$ is the number of bits on which they disagree, so $h(x,y)=w(x\oplus y).$
    \item A subset of $\{0,1\}^n$ is said to be a \textbf{code}. An element of a code is a \textbf{codeword}. A code is said to be \textbf{linear} if it is closed under $\oplus$. It is easy to see that a vector space is formed by a linear code over the field $\{0,1\}$.
    \item If $n$ is even, then we will call $x\in\{0,1\}^n$ \textbf{semi-weight} if $w(x)=\frac{n}{2}$.
    \item The \textbf{complement} $\overline{x}$ of $x\in\{0,1\}^n$ replaces each $0$ in $x$ with $1$, and each $1$ in $x$ with $0$. For example, $\overline{001100}=110011.$
    \item For $x\in\{0,1\}^n$ and $y\in\{0,1\}^m$, denote by $x|y$ the \textbf{concatenation} in $\{0,1\}^{n+m}$ of $x$ to the left of $y$. For example, $101|010=101010.$
\end{enumerate}
\end{definition}

\begin{definition}
For each integer $k$ such that $1\le k\le n,$ let $H_k^n$ be the set of unordered pairs from $\{0,1\}^n$ such that the two elements of the pair have a Hamming distance of $k.$ In particular, we will be focused on $H_1^n,$ which we will call $E_n$. The $E$ in $E_n$ stands for ``edges,'' for a reason that will be explained momentarily.
\end{definition}

It will be helpful to visualize the $n$-dimensional hypercube as the graph whose vertices are the $2^n$ elements of $\{0,1\}^n$ and whose edges are the elements of $E_n.$ Note that we can construct the $(n+1)$-cube recursively by appending a $0$ to the right end of each element of $\{0,1\}^n$ for one copy, then appending a $1$ to the right end of the each element of $\{0,1\}^n$ for another copy, and connecting corresponding elements of the two sets by an edge. We will be seeking functions mapping from $\{0,1\}^n\to \{0,1\}^m$ that leave particular distributions of Hamming distances between outputs corresponding to inputs that are pairs of vertices from edges of the $n$-cube.

\begin{lemma}
The cardinality of $H_k^n$ is $\binom{n}{k}\cdot 2^{n-1}$. As a special case, the cardinality of $E_n=H_1^n$ is $n2^{n-1}$, making it divisible by $n$.
\end{lemma}

\begin{proof}
We will compute the number of \textit{ordered} pairs that have a Hamming distance of $k$ and divide the result by $2.$ Firstly, $\binom{n}{k}$ allows us to choose $k$ out of $n$ indices at which the two strings differ and $n-k$ indices at which they are the same. There are $2^n$ ways to choose the first string, each of which cause the second string to be fixed because we know exactly the indices at which they agree or differ. Finally, we divide by $2$ to account for the fact that $H_k^n$ consists of \textit{unordered} pairs, which yields the formula $$|H_k^n|=\binom{n}{k}\cdot 2^{n}\cdot \frac{1}{2}=\binom{n}{k}\cdot 2^{n-1}.$$
\end{proof}

\begin{definition}
There are three kinds of scattering or diffusive functions that we will explore:
\begin{enumerate}
    \item One notion is that, if a bit is flipped ($0$ to $1$, or $1$ to $0$) in any fixed input, then some half of the bits in the corresponding output flips. Formally, we will say that an injective function $a:\{0,1\}^n\to\{0,1\}^m$ is \textbf{absolutely diffusive} if $m$ is even and
	$$\forall \{x,y\}\in E_n: h(a(x),a(y))= \frac{m}{2}.$$
	\item A second concept is that if a bit is flipped in any fixed input, then, for each bit in the corresponding output, the probability of it flipping is half. Formally, an injective function $p:\{0,1\}^n\to\{0,1\}^m$ is said to be \textbf{probabilistically diffusive} if, for each index $1\le i\le n$, we have
	$$\forall \{x,y\}\in E_n: \frac{1}{n2^{n-1}}\cdot\sum_{\{x,y\}\in E_n}{\pi_i(p(x)\oplus p(y))}= \frac{1}{2},$$ where $\pi_i:\{0,1\}^n\to\mathbb{Z}$ is the projection map that takes the $i^{\text{th}}$ bit of its input and maps $0,1\in\{0,1\}$ to $0,1\in\mathbb{Z}$ respectively.
	\item A third idea is that an injective function $u:\{0,1\}^n \to \{0,1\}^n$ is said to be \textbf{uniformly diffusive} if the distribution of Hamming distances between the outputs of every pair in $E_n$ is uniform on $\{1,2,3,\ldots, n\}.$
\end{enumerate}
\end{definition}

\begin{theorem}
For positive integers $n\equiv 0,1,2,3\pmod{4}$ respectively, the minimum positive integer $m$ for which there exists a absolutely diffusive map $$a_n:\{0,1\}^n\to\{0,1\}^m$$ is $m= n+2,n+1,n,n+1.$ Every output dimension greater than the minimum also allows for absolutely diffusive functions with input dimension $n$.
\end{theorem}

\begin{proof}
For each positive integer $n$, $\{0,1\}^n$ has $2^n$ elements, so in order for an injective function $a_n:\{0,1\}^n\to\{0,1\}^m$ to exist, $m\ge n$. We will show that $m=n$ can be achieved for $n\equiv 2\pmod{4}$. For odd $n$, since we always require even $m$, we need $m\ge n+1$ which we will also prove to be achievable. However we will show that, for $n\equiv 0\pmod{4}$, no permutation of $\{0,1\}^n$ is absolutely diffusive, so we will instead find a map to $\{0,1\}^{n+2}$. For contradiction, suppose $m=n$ is achievable in the case of $4\mid n$. Let $$x=x_1\cdots x_n\in\{0,1\}^n.$$ Suppose its non-zero bits are at indices $i_1,\ldots, i_k$. Let $(w_0,w_1,\ldots, w_k)$ be the $(k+1)$-tuple of vectors from $\{0,1\}^n$ where $w_0$ is the $0$-vector and in each other $w_j$ the non-zero bits are precisely those at indices $i_1,\ldots, i_j$. Note that the absolutely diffusive property is preserved under XOR-ing all outputs by the same element of $\{0,1\}^m$ due to the fact that $$(A\oplus C)\oplus (B\oplus C)=A\oplus B.$$ So we may assume without loss of generality that the $0$-vector maps to the $0$-vector. Note that XOR-ing two even weight vectors produces another vector of even weight and that $$A \oplus (A \oplus B) = B.$$ Since $a_n(0)=0$ has even weight, and consecutive elements of $(w_0,w_1,\ldots,w_k)$ differ by exactly one bit, and each $$a_n(w_i)\oplus a_n(w_{i+1})$$ is semi-weight and so even weight, we can produce a chain reaction to find that each $a_n(w_i)$ has even weight. Thus $w_k=x$ (and so every vector) must map to a vector with even weight, of which there are only $2^{n-1}<2^n$. This is the contradiction that we seek.

\bigskip

The case of $n=2$ is trivial as the identity map works so we will assume that $n\ge 3$ now. The space $\{0,1\}^n$ is a linear code which can be generated by the row vectors $(e_{1;n},\ldots, e_{n;n})$ of the $n\times n$ identity matrix. For $k\equiv 0,2\pmod{4}$ respectively, we will soon define $k-1$ and $k$ linearly independent semi-weight vectors $v_{i;k}\in\{0,1\}^k$ and apply them in the following way. Recall that each element of $\{0,1\}^n$ can be uniquely represented as a sum of the $e_{i;n}$. Define
$$a_n\left(\bigoplus_{i=1}^{n}{c_i\cdot e_{i;n}}\right)= \bigoplus_{i=1}^{n}{c_i\cdot v_{i;m}},$$
where $m=n+2, n+1, n, n+1$ respectively for $n\equiv 0,1,2,3\pmod{4}$, and the $c_i$ are from $\{0,1\}$. The map $a_n$ will be injective due to the linear independence of the $v_{i;m}$, and the semi-weight property of the $v_{i;m}$ means that a $1$-bit change in the input causes a change in some half of the bits in the output.

\bigskip

Finally, we define the $v_{i;k}$. The identity matrix has linearly independent rows $(e_{1;k},\ldots,e_{k;k})$, so the elementary operation of adding one row to another in this matrix with our vector addition $\oplus$ preserves linear independence. For even integers $k\ge 4$, we define
$$v_{i;k}=
\begin{cases}
	\bigoplus_{j=i}^{i+\frac{k}{2}-1}{e_{j;k}} & \text{ if } 1\le i\le \frac{k}{2} \\
	e_{i;k}\oplus v_{\frac{k}{2};k}\oplus e_{k;k} & \text{ if } \frac{k}{2}+1\le i\le k-1 \\
	e_{k;k}\oplus v_{1;k}\oplus v_{\frac{k+2}{4};k} & \text{ if } i=k.
\end{cases}$$
The final vector is only for $k\equiv 2\pmod{4}$. Since the $v_{i;k}$ are all semi-weight, the construction is complete. It is evident that the dimension of the output space can be chosen to be any even integer greater than $m$, if desired, by using some of the $v_{i;k}$ from the higher dimension.
\end{proof}

\begin{example}
For $n=6$, our absolutely diffusive map $a_6$ has basis
$$\begin{bmatrix}
v_{1;6} \\ 
v_{2;6} \\ 
v_{3;6} \\ 
v_{4;6} \\ 
v_{5;6} \\ 
v_{6;6} 
\end{bmatrix} 
= 
\begin{bmatrix}
1 & 1 & 1 & 0 & 0 & 0 \\
0 & 1 & 1 & 1 & 0 & 0 \\
0 & 0 & 1 & 1 & 1 & 0 \\
0 & 0 & 1 & 0 & 1 & 1 \\
0 & 0 & 1 & 1 & 0 & 1 \\
1 & 0 & 0 & 1 & 0 & 1
\end{bmatrix}.$$
In the $n\equiv 2\pmod{4}$ case, if the column vectors are used as the $v_{i;k}$ instead of the rows, then for each bit in the output, a change in one of exactly some half of the input bits causes that output bit to flip. Since column rank equals row rank, linear independence of the rows implies that of the columns, causing injectivity. Therefore, this produces a probabilistically diffusive map. Unfortunately, the matrices are not square in the other three cases. Interestingly, we will see that there exists a probabilistically diffusive permutation of every input space $\{0,1\}^n$ for integers $n\ge 2$.
\end{example}

\begin{definition}
For the next theorem, we will need the following functions.
\begin{enumerate}
	\item For each bit string $x\in\{0,1\}^n$ and integer $n\ge 2$, let $\alpha(x)$ denote the rightmost bit of $x$ as an element of $\{0,1\}$. Similarly, let $\beta(x)$ denote the string in $\{0,1\}^{n-1}$ when the rightmost bit of $x$ is removed.
	\item Let $\tau:\{0,1\}^n\to \{0,1\}^n$ denote the function which transposes the leftmost two bits. Consequently define $\sigma:\{0,1\}^n\to \{0,1\}^n$ by $\sigma(x)=\tau(x\oplus 1)$, where we have used the generalized definition of $\oplus$. Geometrically, $\sigma$ is a clockwise rotation by a quarter circle.
\end{enumerate}
It is useful to note for later that $\beta\circ\sigma=\sigma\circ\beta$ for integers $n\ge 3$.
\end{definition}

\begin{theorem}
For each integer $n\ge 2$, there exists a probabilistically diffusive function $$p_n:\{0,1\}^n\to \{0,1\}^n$$ and the same is true for every larger output dimension $m\ge n$. Of course, we cannot have the output dimension be less than $n$ due to injectivity.
\end{theorem}

\begin{proof}
It is straightforward to verify that for $n=2$, the identity function
\begin{align*}
    p_2:\{0,1\}^2&\to \{0,1\}^2\\
    x&\mapsto x
\end{align*}
is probabilistically diffusive. We recursively define $p_n$ for integers $n\ge 3$ as

$$p_n(x)=\begin{cases}
	(p_{n-1}\circ \beta)(x)|(\alpha\circ p_{n-1}\circ \beta)(x) & \text{ if } \alpha(x) =0\\
	(p_{n-1}\circ \beta\circ\sigma)(x)|\overline{(\alpha\circ p_{n-1}\circ \beta\circ \sigma)(x)} &\text{ if } \alpha(x) =1.
\end{cases}$$

We must prove that $p_n$ is bijective and satisfies the probabilistically diffusive property that for every index $1\le i\le n$, $$\sum_{\{x,y\}\in E_n}{\pi_i(p_n(x)\oplus p_n(y))}=n2^{n-2}. $$
Both will be proven by induction. With $p_2$ as the basis established, suppose that for some integer $n-1\ge 2$, it holds that $p_{n-1}$ is bijective and satisfies the probabilistically diffusive equation above.

We can easily see that $\sigma$ is a bijection of $\{0,1\}^n$ because, for each $x\in\{0,1\}^{n-2}$, it cycles: $$00|x\to 01|x\to 11|x\to 10|x\to 00|x \to \cdots.$$
This property is also important in the latter half of the proof. Then, by the bijectivity of $p_{n-1}$,
\begin{align*}
	p_n(\{0,1\}^n) =& \{ p_{n-1}(z)|(\alpha\circ p_{n-1})(z) : z\in\{0,1\}^{n-1} \}\\
	&\cup \{ (p_{n-1}\circ\sigma)(z)|\overline{(\alpha\circ p_{n-1}\circ \sigma)(z)} : z\in\{0,1\}^{n-1} \}\\
	=& \{ p_{n-1}(z)|(\alpha\circ p_{n-1})(z) : z\in\{0,1\}^{n-1} \}\\
	&\cup \{ p_{n-1}(z)|\overline{(\alpha\circ p_{n-1})(z)} : z\in\{0,1\}^{n-1} \}\\
	=& \{ z|0 : z\in\{0,1\}^{n-1} \}\cup \{ z|1 : y\in\{0,1\}^{n-1} \}\\
	=& \{0,1\}^n,
\end{align*}
proving bijectivity of $p_n$.

For the probabilistically diffusive property, first note that for each index $1\le i\le n$,
\begin{align*}
\sum_{\{x,y\}\in E_n}{\pi_i(p_n(x)\oplus p_n(y))}=&
\sum_{\phi=0,1}\sum_{\{x,y\}\in E_{n-1}}{\pi_i(p_n(x|\phi)\oplus p_n(y|\phi))}\\
&+
\sum_{x\in\{0,1\}^{n-1}}{\pi_i(p_n(x|0)\oplus p_n(x|1))}.
\end{align*}
From left to right, name these four sums $c_i, r_i, s_i, t_i$ ($r_i$ for $\phi=0$ and $s_i$ for $\phi =1$). We have $r_n=r_{n-1}$ since the rightmost bit in each of the two components of $$p_n(x|0)\oplus p_n(y|0)$$ is simply a copy of the second-from-rightmost bit of that component by the first line of the definition of $p_n$. Similarly $s_n=s_{n-1}$ as the rightmost bit in each of the two components of $$p_n(x|1)\oplus p_n(y|1)$$  is the complement of the second-from-rightmost bit of that component by the second line of the definition of $p_n$, and using $$\overline{A}\oplus \overline{B}=A\oplus B.$$ So for the $r_i$ and $s_i$, we may assume without loss of generality that $1\le i\le n-1$. Then
\begin{align*}
    r_i &= \sum_{\{x,y\}\in E_{n-1}}{\pi_i(p_n(x|0)\oplus p_n(y|0))}\\
    &=\sum_{\{x,y\}\in E_{n-1}}{\pi_i(p_{n-1}(x)\oplus p_{n-1}(y))}\\
    &=(n-1)2^{n-3},
\end{align*}
by the induction hypothesis. Since $\sigma$ is bijective on $\{x|0:x\in\{0,1\}^{n-1}\}$ and we have the following identities for all $x,y\in\{0,1\}^{n-1}$
\begin{align*}
	p_n(x|1)\oplus p_n(y|1)&=(p_n\circ\sigma)(x|0)\oplus (p_n\circ\sigma)(y|0), \\
	h(\sigma(x),\sigma(y))&=h(x,y)
\end{align*}
immediately from definitions, it follows that
\begin{align*}
	s_i &= \sum_{\{x,y\}\in E_{n-1}}{\pi_i(p_n(x|1)\oplus p_n(y|1))}\\
	&=\sum_{\{x,y\}\in E_{n-1}}{\pi_i((p_n\circ\sigma)(x|0)\oplus (p_n\circ\sigma)(y|0))}\\
	&= \sum_{\{x,y\}\in E_{n-1}}{\pi_i(p_n(x|0)\oplus p_n(y|0))}\\
	&= r_i\\
	&= (n-1)2^{n-3}.
\end{align*}
Using the fact that $A\oplus \overline{B}=\overline{A\oplus B}$,
\begin{align*}
	t_n &= \sum_{x\in\{0,1\}^{n-1}}{\pi_n(p_n(x|0)\oplus p_n(x|1))}\\
	&= \sum_{x\in\{0,1\}^{n-1}}{\pi_n((\alpha\circ p_{n-1})(x)\oplus \overline{(\alpha\circ p_{n-1}\circ\sigma)(x)})}\\
	&= \sum_{x\in\{0,1\}^{n-1}}{\pi_n(\overline{p_{n-1}(x)\oplus (p_{n-1}\circ \sigma)(x)})}\\
	&= 2^{n-1} - t_{n-1},
\end{align*}
since there are $2^{n-1}$ terms in the sum and the above is the complement of $t_{n-1}$ (see the sum for $t_i$ below). We will prove that $$t_1=t_2=\cdots =t_{n-1}=2^{n-2},$$ which will result in $$t_n=2^{n-1}-t_{n-1}=2^{n-1}-2^{n-2}=2^{n-2},$$ so assume without loss of generality that $1\le i\le {n-1}$. Then
\begin{align*}
t_i &= \sum_{x\in\{0,1\}^{n-1}}{\pi_i(p_n(x|0)\oplus p_n(x|1))}\\
&=\sum_{x\in\{0,1\}^{n-1}}{\pi_i(p_{n-1}(x)\oplus (p_{n-1}\oplus\sigma)(x))}.
\end{align*}
Recalling how the permutation representation of $\sigma$ is the product of all $4$-cycles $$(00|x,01|x,11|x,10|x),$$ the above sum can be written as the sum over all $z\in\{0,1\}^{n-3}$ of the quadruple sum
\begin{align*}
	&\pi_i(p_{n-1}(00|z)\oplus p_{n-1}(01|z))\\
	+& \pi_i(p_{n-1}(01|z)\oplus p_{n-1}(11|z))\\
	+& \pi_i(p_{n-1}(11|z)\oplus p_{n-1}(10|z))\\
	+& \pi_i(p_{n-1}(10|z)\oplus p_{n-1}(00|z)).
\end{align*}
It is a standard induction argument over $n\ge 3$ (with casework on $i<n$ vs. $i=n$, and casework on $\alpha(z)=0,1$, for a total of $2\cdot 2 =4$ cases) to prove that each such quadruple sum is equal to $2$ for $1\le i\le n-1$. So the sum over all such quadruple sums is $$2\cdot 2^{n-3}=2^{n-2}.$$ Therefore, $$c_i=r_i+s_i+t_i=(n-1)2^{n-3}+(n-1)2^{n-3}+2^{n-2}=n2^{n-2},$$
as desired. Moreover, by appending the rightmost bit of each output to that output some fixed number of times, a probabilistically diffusive injection into any output space of larger dimension  can be constructed.
\end{proof}

\begin{example}
For $n=3$, we have the probabilistically diffusive map $p_3$
$$
\begin{cases}
000\mapsto 000\\
001\mapsto 010\\
010\mapsto 011\\
011\mapsto 110
\end{cases}
\hspace{1mm}\text{and}\hspace{2mm} \begin{cases}
100\mapsto 100\\
101\mapsto 001\\
110\mapsto 111\\
111\mapsto 101
\end{cases}.
$$
This was constructed using the recursive definition of our probabilistically diffusive maps $p_n$, which begin with the identity function for $n=2$.
\end{example}

\begin{theorem}
For each positive integer $n,$ there exists a uniformly diffusive function $$u_n:\{0,1\}^n \to \{0,1\}^n$$ such that the distribution of Hamming distances between the outputs of every pair in $E_n$ is uniform on $\{1,2,3,\ldots, n\}.$ It follows that the output dimension can be taken to be any integer $m\ge n$ because we can take $u_n$ and append $m-n$ copies of $0$ to the right end of each output.
\end{theorem}

\begin{proof}
It is plausible that such a function exists because we proved earlier that $|E_n|=n2^{n-1},$ which is divisible by $n,$ and $n$ is the number of values in the set of Hamming distances $\{1,2,3,\ldots, n\}$ that we are targeting.

\bigskip

There are two ``canonical'' copies of $\{0,1\}^n$ in $\{0,1\}^{n+1}$ that we can construct as follows: append a $0$ to the right end of each element of $\{0,1\}^n$ to get the first copy, and  append a $1$ in the same manner to get the second copy. The idea behind constructing $u_n$ is recursion: if $c\in \{0,1\}^{n+1}$ ends in a $0$ (so $c$ lies in the first copy) then map $c$ to what $c$ without the final $0$ would get mapped to, along with a $0$ at the end of the output; if $c$ lies in the second copy, then map it to the complement of what the corresponding element in the first copy gets mapped to. Formally, we define the first $u_1$ as the identity map
\begin{align*}
    u_1: \{0,1\}^1 &\to \{0,1\}^1\\
    0 &\mapsto 0, \text{ and } 1 \mapsto 1,
\end{align*}
and then recursively define $u_{n+1}$ for integers $n\ge 1$ as the following, for all $x\in\{0,1\}^n$:
\begin{align*}
    u_{n+1}(x|0) &= u_n (x)|0,\\
    u_{n+1}(x|1) &= \overline{u_n (x)|0}.
\end{align*}
It is clear that the identity map $u_1$ has the uniformity property, and it remains to be shown that so does $u_n$ for $n\ge 2.$ We will prove that $u_{n+1}$ has the uniformity property based on the uniformity property of $u_n,$ so formally, this is a proof by induction. Basically, we want a uniform distribution of the values $\{1,2,3,\ldots,n,n+1\}$ on the edges of the $(n+1)$-cube after assigning the output values in the stated way. We will show by induction that there are $2^n$ copies of each possible Hamming distance. In the first copy of $\{0,1\}^n$ in $\{0,1\}^{n+1},$ we have this uniformity property by the induction hypothesis. The second copy inherits the same the uniformity property because
\begin{align*}
    h(u_{n+1}(x|1),u_{n+1}(y|1)) &= h\left(\overline{u_{n+1}(x|0)},\overline{u_{n+1}(y|0)}\right)\\
    &= h(u_{n+1}(x|0),u_{n+1}(y|0))\\
    &= h(u_n(x),u_n(y)).
\end{align*}
So each of $\{1,2,3,\ldots, n\}$ is achieved $2\cdot 2^{n-1} = 2^n$ times. Finally, when attaching corresponding elements between the first and second copies of $\{0,1\}^n$ in $\{0,1\}^{n+1},$ every pair differs on every index, so $n+1$ is achieved $2^n$ times as well.
    
\bigskip
    
Every element of $\{0,1\}^{n+1}$ actually gets hit as an output by $u_{n+1}$ because, by induction, we are taking $\{0,1\}^n$ and appending a  $0$ to the right end of each element to get the outputs of the first copy, and then inverting each element to produce the outputs of the second copy.
\end{proof}

\begin{example}
For $n=3$, we have the uniformly diffusive map $u_3$
$$
\begin{cases}
000\mapsto 000\\
001\mapsto 111\\
010\mapsto 110\\
011\mapsto 001
\end{cases}
\hspace{1mm}\text{and}\hspace{2mm} \begin{cases}
100\mapsto 100\\
101\mapsto 011\\
110\mapsto 010\\
111\mapsto 101
\end{cases}.
$$
This was constructed using the recursive definition of the uniformly diffusive maps $u_n$, which begin with the identity function for $n=1$.
\end{example}

There are a number of questions that arise naturally from our definitions and results.
\begin{enumerate}
    \item If a diffusive function of a particular kind (absolute, probabilistic, or uniform) exists and maps from $\{0,1\}^n$ to $\{0,1\}^m$, then there actually exist at least $2^m$ such functions. This is thanks to the preservation of each of the three kinds of diffusion under $\oplus$ of all outputs by some fixed element of $\{0,1\}^m$. One question asks for upper bounds and a stronger lower bound on the number of possible diffusive functions on $\{0,1\}^n$ for each output dimension and for each of the three kinds of diffusion.
    \item We could attempt to construct maps which are combinations of being absolutely, probabilistically, and uniformly diffusive.
    \item Another idea is to modify the definitions of our forms of diffusion to be over pairs of inputs that are elements of $$H_k^n \text{ or } \bigcup_{i=1}^{k}{H_i^n}$$ for $2\le k\le n$ instead of $E_n$. In this framework, we have constructed optimal functions for $k=1$. Bounds on the minimal output dimensions for other small $k\ge 2$ and the construction of these near-optimal maps would be an accomplishment. Presumably, research into the graph structure of $H_k^n$ will be useful. It is helpful for visualization to note that $H_k^n$ is the set of unordered pairs of vertices of the $n$-cube such that the minimal path between the two vertices traverses $k$ edges. Also, $\bigcup_{i=1}^{n}{H_i^n}$ is the complete graph on $2^n$ vertices, so starting further work by looking at that extreme end could be fruitful.
\end{enumerate}

\begin{center}
    Acknowledgements
\end{center}

The author thanks the ROP cryptography research group of 2012-2013 at the University of Toronto for their support and camaraderie.

\begin{center}
    References
\end{center}

\begin{enumerate}
    \item C. Shannon, (1949) \textit{Communication Theory of Secrecy Systems}, Bell System Technical Journal, Vol. 28, pp. 656-715
\end{enumerate}

\end{document}